\documentclass[showpacs,amsmath,pra]{revtex4}
\usepackage{graphicx}
\usepackage{dcolumn}
\usepackage{bm}

\begin{document}

\title{The average detection probability of the quantum dialogue protocol under the disturbance attack is 3/4}
\author{Nguyen Ba An}
\email{nbaan@kias.re.kr}
\affiliation{School of Computational Sciences, Korea Institute for Advanced Study,\\
 207-43 Cheongryangni 2-dong, Dongdaemun-gu, Seoul 130-722, Republic of  Korea }

\begin{abstract}
We prove explicitly that the detection probability of the disturbance attack in the 
recently proposed quantum dialogue protocol is 3/4 in average. The purpose is not only to reply a comment 
but also to provide a deeper understanding of a kind of tampering in an unauthorized communication.
\end{abstract}
\pacs{03.67.Hk}

\maketitle

Of late Cai made a comment \cite{cai} on our protocol \cite{an} that enables two authorized 
parties to securely exchange their messages in a deterministic and direct manner resembling 
a dialogue, hence the name ``quantum dialogue" protocol (a similar idea with less capacity 
was touched upon earlier in \cite{zhang1} where no full sets of steps necessary enough to securely 
process the protocol were shown; 
see also \cite{zhang2} for a different possible scheme). Cai remarked that the detection probability 
$d$ of the disturbance attack per control mode run is 1/2 but not 3/4 as said in \cite{an}. 

First, if Cai's notification were right, it does not so much influence the quantum dialogue protocol  
which remains still {\it asymptotically} secure.  Second, however, in average $d$ should be 3/4 
indeed, instead of 1/2, as will be shown explicitly now. To make it rigorous and explicit as 
much as possible we shall not be shy 
to expose the mathematical formulation in full detail in what follows.

At the beginning, i.e. in the preparation stage, Bob is supposed to be provided with 
a large enough number of maximally two-qubit entangled states (Einstein-Podolsky-Rosen pairs)
 \begin{equation}
\left| \Psi_{0,0}\right\rangle _{ht}=\frac{1}{\sqrt{2}}\left( \left|
0 \right\rangle _{h}\left| 1\right\rangle
_{t}+\left| 1 \right\rangle _{h}\left| 0 \right\rangle _{t}\right).\label{e0}
\end{equation}
Bob encodes his two secrete bits $k,l \in \{0,1\}$ by applying $C^t_{k,l}$ ($C^t_{0,0}=\widehat{1}, 
C^t_{0,1}=\sigma^t_x, C^t_{1,0}=\sigma^t_y, C^t_{1,1}=\sigma^t_z$, and $\sigma^t_{x,y,z}$ the Pauli matrices) 
on qubit $t$ to have the state
\begin{equation}
\left| \Psi_{k,l}\right\rangle _{ht}=\frac{1}{\sqrt{2}}\left( \left|
0 \right\rangle _{h}C^t_{k,l}\left| 1\right\rangle
_{t}+\left| 1 \right\rangle _{h}C^t_{k,l}\left| 0 \right\rangle _{t}\right),\label{e1}
\end{equation}
and sends qubit $t$ to Alice. One way to disturb the communication is that Eve, the eavesdropper, 
measures the qubit $t$ on the $B\rightarrow A$ route (similar results will be obtained if Eve's measurement 
is to be made on the $A\rightarrow B$ route) \cite{caiprl}. Eve's measurement outcome is that 
either 
\begin{description}
\item[(a)] she finds the state $ \left| 1 \right\rangle _{t}$ or
\item[(b)] she finds the state $\left| 0 \right\rangle _{t}$,
\end{description}
with a probability of 50\% 
each. Then Eve sends on the measured qubit $t$ to Alice.   In case (a) the Alice-Bob pair 
state collapses into
\begin{equation}
 \left| 0 \right\rangle _{h}C^t_{k,l}\left| 1\right\rangle _{t},\label{e2}
\end{equation}
whereas in case (b) it is
\begin{equation}
 \left| 1 \right\rangle _{h}C^t_{k,l}\left| 0\right\rangle _{t}.\label{e3}
\end{equation}
Alice being unaware of Eve's tampering encodes her two confidential bits $i,j\in \{0,1\}$ by applying 
$C^t_{i,j}$ on the qubit $t$ Eve has just sent her, then returns it back to Bob. In case (a) the Alice-Bob pair state becomes
\begin{equation}
 \left| \Phi^{(a)}\right\rangle_{ht}=\left| 0 \right\rangle _{h}C^t_{i,j}C^t_{k,l}\left| 1\right\rangle _{t}
=\left|0\right\rangle_hC^t_{i\oplus k,j\oplus l}\left|1\right\rangle_t,\label{e4}
\end{equation}
whereas in case (b) it is
\begin{equation}
 \left| \Phi^{(b)}\right\rangle_{ht}=\left| 1 \right\rangle _{h}C^t_{i,j}C^t_{k,l}\left| 0\right\rangle _{t}
=\left|1\right\rangle_hC^t_{i\oplus k,j\oplus l}\left|0\right\rangle_t.\label{e5}
\end{equation}
In Eqs. (\ref{e4}) and (\ref{e5}) the $\oplus $ stands for an addition mod 2 and we have omitted the unimportant 
phase factor $\phi_{i,j,k,l}$ for simplicity (see \cite{an} for its explicit expression). 

Now we observe the following identities 
 \begin{equation}
C^t_{a,b}\left| 1\right\rangle _{t}\equiv (-1)^a \delta_{a\oplus b,0}\left|1\right\rangle_t 
+(i)^a\delta_{a\oplus b,1}\left|0\right\rangle_t\label{e6}
\end{equation}
and
 \begin{equation}
C^t_{a,b}\left| 0\right\rangle _{t}\equiv  \delta_{a\oplus b,0}\left| 0\right\rangle_t 
+(-i)^a\delta_{a\oplus b,1}\left| 1\right\rangle_t\label{e7}
\end{equation}
which can directly be verified for any $a,b\in \{0,1\}$. In Eqs. (\ref{e6}) and (\ref{e7})  
$\delta_{m,n}$ is the Kroneker symbol.  
Using the identities (\ref{e6}) and (\ref{e7}) in Eqs. (\ref{e4}) and (\ref{e5}), respectively, 
yields
 \begin{equation}
\left| \Phi^{(a)}\right\rangle_{ht}= (-1)^{i\oplus k}  \delta_{J,0}\left| 01\right\rangle_{ht} 
+(i)^{i\oplus k}\delta_{J,1}\left| 00\right\rangle_{ht}\label{e8}
\end{equation}
and
 \begin{equation}
\left| \Phi^{(b)}\right\rangle_{ht}=  \delta_{J,0}\left|1 0\right\rangle_{ht} 
+(-i)^{i\oplus k}\delta_{J,1}\left| 11\right\rangle_{ht}\label{e9}
\end{equation}
where $J\equiv i\oplus k\oplus j\oplus l$. 

Another universal (i.e., valid for any $\mu,\nu\in \{0,1\}$) identity which can also be 
justified by direct inspection is 
 \begin{equation}
\left| \mu\nu\right\rangle_{ht}=\frac{1}{\sqrt{2}}
\left(   \left|\Psi_{0,\mu\oplus\nu}\right\rangle_{ht}
+ (-1)^\mu \left|\Psi_{1,\mu\oplus\nu}\right\rangle_{ht}
 \right). \label{e10}
\end{equation}
Use of the identity (\ref{e10}) in Eqs. (\ref{e8}) and (\ref{e9}) re-expresses the latter in the Bell basis as
 \begin{eqnarray}
\left| \Phi^{(a)}\right\rangle_{ht}&=&
\frac{1}{\sqrt{2}} (-1)^{i\oplus k} \delta_{J,0}
\left(   \left|\Psi_{0,1}\right\rangle_{ht}
+  \left|\Psi_{1,1}\right\rangle_{ht}\right)\nonumber\\
& +& \frac{1}{\sqrt{2}} (i)^{i\oplus k} \delta_{J,1}
\left(   \left|\Psi_{0,0}\right\rangle_{ht}
+  \left|\Psi_{1,0}\right\rangle_{ht}
 \right) \label{e11}
\end{eqnarray}
and
 \begin{eqnarray}
\left| \Phi^{(b)}\right\rangle_{ht}&=&
\frac{1}{\sqrt{2}}  \delta_{J,0}
\left(   \left|\Psi_{0,1}\right\rangle_{ht}
-  \left|\Psi_{1,1}\right\rangle_{ht}\right)\nonumber\\
& +& \frac{1}{\sqrt{2}} (-i)^{i\oplus k} \delta_{J,1}
\left(   \left|\Psi_{0,0}\right\rangle_{ht}
-  \left|\Psi_{1,0}\right\rangle_{ht}
 \right). \label{e12}
\end{eqnarray}

The above formulae (\ref{e11}) and (\ref{e12}) allow us to calculate 
the average detection probability. Namely, since $i,j,k,l\in \{0,1\}$ there are four possibilities in all:
\begin{description}
\item[(i)] $\{i\oplus k=0, j\oplus l=0\}$ $\Rightarrow J=0$,
\item[(ii)] $\{i\oplus k=0, j\oplus l=1\}$ $\Rightarrow J=1$,
\item[(iii)] $\{i\oplus k=1, j\oplus l=0\}$ $\Rightarrow J=1$,
\item[(iv)] $\{i\oplus k=1, j\oplus l=1\}$ $\Rightarrow J=0$.
\end{description}
We first deal with case (a) employing Eq. (\ref{e11}).  
In case (a,i) the detection probability is $d_{(a,i)}=1$ since Bob never finds state 
$\left|\Psi_{i\oplus k,j\oplus l}\right\rangle_{ht}=\left|\Psi_{0,0}\right\rangle_{ht}$ as he should 
in Eve's absence. In case (a,ii)  the detection probability is $d_{(a,ii)}=1$ also since Bob never finds state 
$\left|\Psi_{i\oplus k,j\oplus l}\right\rangle_{ht}=\left|\Psi_{0,1}\right\rangle_{ht}$ as he should 
in Eve's absence. In case (a,iii) the detection probability is $d_{(a,iii)}=1/2$ since there is a 50\% probability that Bob finds state 
$\left|\Psi_{i\oplus k,j\oplus l}\right\rangle_{ht}=\left|\Psi_{0,0}\right\rangle_{ht}$ 
indicating Eve's presence. In this case Bob should find state $\left|\Psi_{i\oplus k,j\oplus l}\right\rangle_{ht}=\left|\Psi_{1,0}\right\rangle_{ht}$ 
if there is no eavesdropping, but that just as well occurs with a probability of 50\%. 
Finally,  in case (a,iv) the detection probability is $d_{(a,iv)}=1/2$ also since there is a 50\% probability that Bob 
finds state $\left|\Psi_{i\oplus k,j\oplus l}\right\rangle_{ht}=\left|\Psi_{0,1}\right\rangle_{ht}$ 
indicating Eve's presence. In this case Bob should find state $\left|\Psi_{i\oplus k,j\oplus l}\right\rangle_{ht}=\left|\Psi_{1,1}\right\rangle_{ht}$ 
if there is no eavesdropping, 
but that occurs only with a probability of 50\%.  With all the four above possible situations taken into account in case (a) the average detection probability is calculated as 
\begin{equation}
d_{(a)}=\frac{1}{4}\sum_{q=i}^{iv}d_{(a,q)}=\frac{1}{4}\left( 1+1+\frac{1}{2}+\frac{1}{2}\right)=\frac{3}{4}.
\end{equation}
 Similarly, employing Eq. (\ref{e12}) for case (b) we arrive at the results 
$d_{(b,i)}=d_{(b,ii)}=1$ and $d_{(b,iii)}=d_{(b,iv)}=1/2$ yielding again $d_{(b)}=3/4.$ 
Hence, no matter which measurement outcome Eve gets the  average detection probability 
under the disturbance attack by measuring qubit $t$ {\it en route} is $d=3/4.$

Another way for Eve to disturb information is to apply a $C^t_{u,v}$  on qubit $t$ when it travels in between Alice and Bob. 
In the original (one-way) ping-pong protocol \cite{pingpong}, where control mode and message mode are distinguishable, 
Eve can surely avoid all the runs in control mode. Then the intended message becomes absolutely meaningless 
if Eve randomly applies  $C^t_{0,0}$ or $C^t_{1,1}$ on qubit $t$ in every message mode run, as mentioned in \cite{an}. 
Yet, in the (two-way) quantum dialogue protocol presented in \cite{an} the control mode 
was modified so that Eve can by no means distinguish between runs in control mode and in message mode. 
Thus, she cannot avoid a control mode run and  the quantum dialogue protocol in \cite{an} is {\it immune} from 
such a kind of disturbing trick. Nevertheless, if the obstinate Eve still wishes to do that trick, 
the Alice-Bob pair state when qubit $t$ came back to Bob after 
Alice's encoding and after Eve had at random applied an operator $C^t_{u,v}$ (with $u,v\in \{0,1\}$) 
is $\left|\Psi_{i\oplus k\oplus u,j\oplus l\oplus v}\right\rangle_{ht},$ up 
to a global phase factor. Unless $u=v=0$, Eve is inevitably disclosed after Bob's Bell measurement.  
Obviously, the   probability of detecting Eve is again 3/4.

This work is funded by the KIAS R\&D grant No. 03-0149-002.

\end{document}